\documentclass[aps,prl,twocolumn,groupedaddress,amsmath,amssymb]{revtex4}

\usepackage{graphicx}
\usepackage{color}
\usepackage{comment}
\usepackage{bm}
\usepackage{latexsym}

\usepackage[symbol]{footmisc}
\usepackage{hyperref}

\usepackage[T1]{fontenc}
\usepackage[latin9]{inputenc}
\usepackage{esint}

\begin{document}

\title{Adiabatic Thermal Radiation Pumps for Thermal Photonics}
\author{Huanan Li$^{1,2}$\footnotemark[1], Lucas J. Fern\'andez-Alc\'azar$^{1}$\footnotemark[1], Fred Ellis$^1$, Boris Shapiro$^3$, Tsampikos Kottos$^1$} 
\affiliation{$^1$Wave Transport in Complex Systems Lab, Department of Physics, Wesleyan University, Middletown, CT-06459, USA\\
$^2$Photonics Initiative, Advanced Science Research Center, CUNY, NY 10031, USA\\
$^3$Technion - Israel Institute of Technology, Technion City, Haifa 32000, Israel}
\date{\today}

\begin{abstract}
We control the direction and magnitude of thermal radiation, between two bodies at equal temperature (in thermal equilibrium), by invoking the 
concept of adiabatic pumping. Specifically, within a resonant near-field electromagnetic heat transfer framework, we utilize an {\it instantaneous} 
scattering matrix approach to unveil the critical role of wave interference in radiative heat transfer. We find that appropriately designed adiabatic 
pumping cycling near diabolic singularities can dramatically enhance the efficiency of the directional energy transfer. We confirm our results using 
a realistic electronic circuit set-up.  
\end{abstract}

\maketitle

\footnotetext[1]{These authors contributed equally to the results of this paper}

{\it Introduction --} Understanding the rules that dictate thermal radiation and the development of novel schemes that allow us to tame its flow, has 
offered over the years an exciting arena of research \cite{VP07,BLID11,HSM11,F17}. From one side, there are fundamental challenges associated 
with basic constraints that need to be understood in order to manage thermal radiation \cite{F17,MZF17,ZF14,SWL98,HSA16}. At their core are 
questions associated with the importance of thermal electromagnetic fluctuations and their implications in directional thermal radiation. On the other 
hand, there is a wide range of applications  that can benefit from advances in thermal radiation management. In fact, in close connection with the 
rapid developments that we are witnessing in the field of nanophotonics, the subfield of thermal photonics has emerged \cite{RIBCJSJ11,AB14,
SNC09,RSJVCCG09,OQWALMRTW11,KHZMFH12,GGZFL14,SLLNS15,Kim15,SGSTFFFVCRM15,GZFL16,STFGRM16} and promises to revolutionize 
modern energy technologies. Examples include thermophotovoltaics \cite{LMF03,NC03,BZF09,IJJCS12}, thermal imaging \cite{WFCGLJMCG06,
KHPBRH05}, thermal circuits \cite{OLF10,BF11,ZOF12}, and radiative cooling \cite{GOPFL12}.

\begin{figure}
\includegraphics[width=1\columnwidth,keepaspectratio,clip]{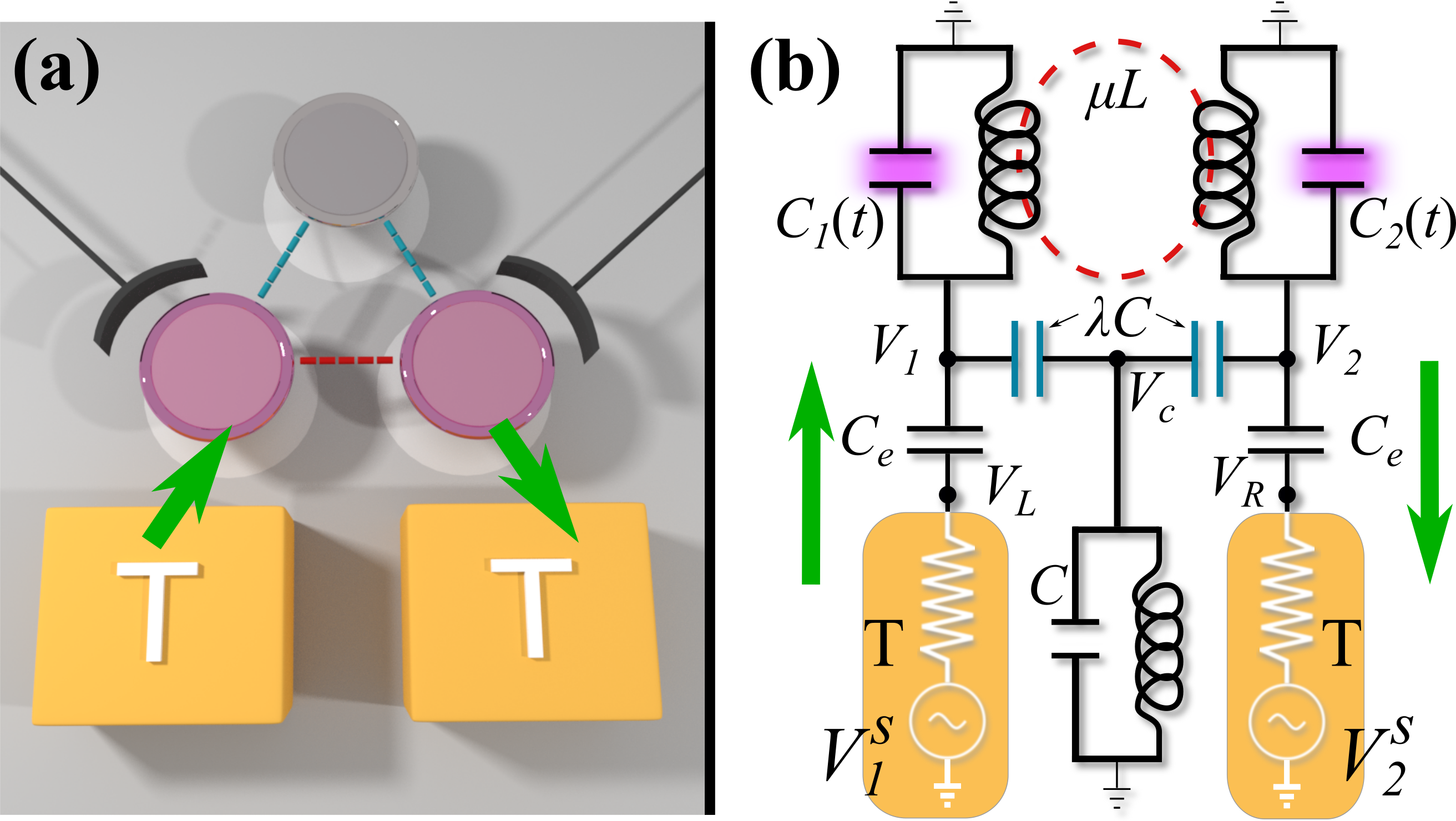}
\caption{Proposed implementations of our thermal radiation pumping scheme: (a) A nano-photonic structure consisting of three single-mode resonators. 
The resonant frequencies of the first and third resonator (purple colors) are periodically modulated via a weak adiabatic modulation of the permittivities 
of the resonators. The system is in contact with two independent baths at the same temperature $T$. (b) A circuit consisting of three LC resonators. Two 
of these resonators are modulated via their (purple) capacitances. The circuit is coupled capacitively to two artificial reservoirs at the same temperature 
$T$. The reservoirs are implemented by synthesized noise sources generating random voltages $V_{1,2}^s$ with a prescribed spectral distribution. The 
positive direction for the pumped flow is chosen to match the green arrows.}
\label{fig1}
\end{figure}

In this paper we propose to manipulate thermal radiation by introducing the concept of an adiabatic thermal radiation pump. The pump operates 
between two reservoirs that are maintained at the same temperature - as opposed to common approach where heat flow requires a temperature 
gradient. A possible setup of a thermal radiation pump is depicted in Fig. \ref{fig1}a. By slow periodic modulation of the eigenfrequencies of two resonators, 
coupled to two separate reservoirs at equal temperature $T$, one can transfer thermal radiation from one reservoir to the other (see green arrows). This process 
of creating a directional radiation flow may be termed {\it adiabatic thermal radiation pumping}. The amount of heat pumped in one cycle depends 
on the details of the modulation process. In particular, we show that the existence of a diabolic point (i.e. an exact degeneracy) in the spectrum of 
the system's Hamiltonian leads to a dramatic enhancement of the effect, for an appropriately chosen modulation cycle. Our theoretical results 
are based on a coupled-mode-theory (CMT) approach to resonant thermal radiation and are backed up by detailed numerical simulations using  
realistic circuit set-ups, see Fig. \ref{fig1}b. Our approach unveils the importance of wave interference in radiative heat transfer by connecting 
the pumped thermal radiative current with the instantaneous reflection phase. This connection opens up new possibilities in the field of thermal 
photonics. Our concept of the adiabatic thermal radiation pumping is inspired by adiabatic charge pumping in  condensed matter, where a DC current in response to a slowly varying time-periodic potential has been proposed \cite{T83,AG99,Levinson2002,Brouwer1998,AEGS01,C03} 
and experimentally demonstrated \cite{SMCG99}.

{\it CMT Modeling of Thermal Radiation--}
We consider a photonic circuit supporting a finite number of resonant modes $N_s$ described by a Hermitian Hamiltonian $H_0$. The system is in contact with two separate heat baths at constant temerature (see Fig. \ref{fig1}a where $N_s=3$) via leads with  the system-lead coupling described by an operator ${\hat W}$.

 At thermal equilibrium, the two baths generate photons at frequency $\omega$ with mean 
number $\Theta_T(\omega)=\left(e^{\hbar\omega\over k_BT}-1\right)^{-1}$ given by the Bose-Einstein statistics. The radiative thermal energy 
exchange between the two heat baths can be studied using a time-dependent CMT \cite{ZSOFSL13,KJ15}
\begin{eqnarray}
\label{CMT}
\imath{d\over dt} \Psi &=&H_{\rm eff} \Psi + \imath {\hat W} \theta^{(+)}; \quad H_{\rm eff}=\left(H_0+\Lambda -{\imath\over 2} {\hat W}{\hat W}^{T}\right)\nonumber\\
\theta^{(-)} &= &{\hat W}^T\Psi -\theta^{(+)}
\end{eqnarray} 
where $\Psi =\left(\psi_1,\psi_2,\cdots,\psi_{N_s}\right)^T$ describes the modal amplitude of the field and it is normalized in a way that $|\psi_s|^2$ 
represents the energy of the $s^{th}$ mode. The variables $\theta_n^{(\pm)}(\omega)$ (frequency domain of $\theta^{\pm}$ in Eq. (\ref{CMT})) 
indicate the flux-amplitudes of the incoming (+) and outgoing (-) waves from and towards the reservoir $n=1,2$ via leads. At thermal equilibrium 
the incoming flux (from the reservoirs) satisfies the correlation relation \cite{ZSOFSL13}
\begin{equation}
\label{bath}
\langle\theta_n^{(+)}(\omega)
(\theta_m^{(+)}(\omega'))^*\rangle={\hbar\omega\over 2\pi}\Theta_T(\omega)\delta(\omega-\omega')\delta_{nm}
\end{equation}
and, therefore, the outgoing power from the $n^{th}$ heat bath is given by the double integral over frequency $\omega$ of this correlation 
function. 

The associated scattering matrix $S$, connecting the outgoing $\theta^{-}(\omega)$ to the incoming $\theta^{+}(\omega)$ waves, 
can be evaluated using  Eq. (\ref{CMT}). We have \cite{Li2017}
\begin{equation}
S =-I_{2}-\imath W^{T}G_{\rm eff}W,\quad G_{\rm eff}=\frac{1}{H_{\rm eff}-\omega I_{N_s}}
\label{St}
\end{equation}   
where $I_{n}$ is the $n\times n$ identity matrix. The matrix $W$, describes the coupling to the leads (in frequency domain), and has 
dimensions $N_s\times 2$. Its elements are $W_{s,n}=\sqrt{v_g} w_n\delta_{sn}$ where $w_n$ are dimensionless coupling strengths, 
and $v_g={\partial\omega(k)\over\partial k}$. Finally, the matrix $\Lambda$ appearing in $H_{\rm eff}$ (see Eq. (\ref{CMT})) is a 
renormalization term due to the coupling of the system with the leads and it is specific to the properties of the leads. 


{\it Thermal Radiation Pumps--}
Next, we consider a system whose Hamiltonian $H_0(u^t,v^t)$ depends on two time-varying independent parameters $(u^t,
v^t)$. We further assume that these parameters are periodically modulated in time with frequency $\Omega$, such that $H_0
\left(t\right)=H_0\left(t+2\pi/\Omega\right)$. During one period of the modulation these parameters form a closed cycle in the 
$(u^t,v^t)$ parameter space. The associated enclosed ``pumping area'' is ${\mathcal A}\equiv\int_{0}^{2\pi/\Omega}dt u^{t}\frac{d v^{t}}
{dt}$. We want to analyze the net radiative energy flux from one bath to another during one pumping circle.

We consider circumstances where the variation of $(u^t,v^t)$ is small such that ${\mathcal A}\rightarrow 0$. We want to 
evaluate the total radiative (time-averaged) thermal energy flux per pumping area $\bar{\mathcal I}$. The latter is:
\begin{equation}
\bar{\mathcal I}\equiv {\Omega\over 2\pi}\int {d\omega\over 2\pi}\hbar \omega \Theta_T(\omega) Q(\omega),\,
Q(\omega)\equiv \lim_{{\mathcal A}\rightarrow0}\frac{\int_{0}^{2\pi\over \Omega}dt\mathcal{I}_{x_0}\left(t,\omega\right)}{{\mathcal A}}
\label{Ibar}
\end{equation}
where $Q(\omega)$ is the radiative energy density (i.e. per area in the parameter space) and $\mathcal{I}_{x_0}(t,\omega)$ 
is the dimensionless (normalized) time-dependent directional net energy current, at some observation cross section 
at $x=x_0$ within the leads. The latter is evaluated under the condition of two uncorrelated counter propagating incoming waves 
of frequency $\omega$ and unit flux.  In the case where $H_0$ is static, the net thermal radiative current $\mathcal{I}$ at each lead is
zero. From Eq. (\ref{Ibar}) it is clear that an 
understanding of $Q\left(\omega\right)$ is essential for the analysis and control of $\bar{\mathcal I}$ \cite{note2}.


{\it Adiabatic Pumping--}
In the adiabatic limit, $\Omega\rightarrow 0$, the study of radiative thermal energy $Q\left(\omega\right)$ boils down to the 
analysis of the {\it instantaneous} scattering matrix $S^{t}$ \cite{Brouwer1998}. The latter is given in terms of Eq. (\ref{St}) 
with the superscript $t$ indicating the parametric dependence of the matrix elements of $S$ at a specific instant $t$ during 
the pumping cycle. It can be generally parametrized in terms 
of three independent parameters: the instantaneous reflectance $R^{t}$, and the instantaneous reflection and transmission 
phases $\alpha^{t}, \varphi^{t}\in\mathcal{R}$ respectively. Specifically we have: 
\begin{align}
S^{t} & =e^{\imath\varphi^{t}}\begin{bmatrix}\sqrt{R^{t}}e^{\imath\alpha^{t}} & \imath\sqrt{1-R^{t}}\\
\imath\sqrt{1-R^{t}} & \sqrt{R^{t}}e^{-\imath\alpha^{t}}
\end{bmatrix},\quad 0\leq R^{t}\leq 1.
\label{param}
\end{align}
Using this parametrization, we write $Q\left(\omega\right)$ as 
\begin{equation}
Q\left(\omega\right) = \lim_{{\mathcal A}\rightarrow0}\frac{1}{{\mathcal A}}\frac{\partial}{\partial\omega}\int_{0}^{2\pi/\Omega}
dtR^{t}\frac{d\alpha^{t}}{dt}=
\frac{\partial}{\partial\omega}\left|\frac{\partial\left(R^t,\alpha^t\right)}{\partial\left(u^t,v^t\right)}\right|
\label{Q_general}
\end{equation}
which applies whenever the period of the driving is larger than the delay time that the ``photons" dwell inside the scatterer. It is important
to stress that Eq. (\ref{Q_general}) allows us to connect wave interference phenomena (imprinted via the reflection phase $\alpha^t$) with
the thermal radiation problem. Furthermore, it opens up new directions in the field of thermal photonics. Using Eqs. (\ref{param},\ref{St}) 
we have that 
\begin{align}
R^{t}=\left|S_{11}^{t}\right|^{2},\; & \frac{d\alpha^{t}}{dt}=\frac{1}{2\imath}\frac{d}{dt}\left(\ln\frac{S_{11}^{t}}{S_{22}^{t}}\right),
\label{Ra}
\end{align}
where the subscripts indicate the matrix elements of $S^{t}$. 

Direct inspection of Eqs. (\ref{param},\ref{Ra}) indicates that the pumped thermal radiation energy will be affected by the proximity
of resonant modes where $R$ and $\alpha$ experience an abrupt change in $\omega$. One would expect that higher order 
spectral singularities, like diabolic points (DP), could lead to a dramatic enhancement of $Q\left(\omega\right)$. {\it We will show 
that their effect in $\bar{\mathcal I}$ is controlled by the position of the adiabatic cycle in the parameter space, 
with respect to such spectral singularities.} 


{\it A prototype CMT model with DPs--}
We consider a prototype system of three coupled resonant modes, that can support a DP degeneracy. The system is described 
by the CMT Hamiltonian
\begin{align}
H_0(t; u^t, v^t) & =\begin{bmatrix}\omega_{0}+v^{t} & -1 & 1\\
-1 & \omega_{0} & -1\\
1 & -1 & \omega_{0}+u^{t}
\end{bmatrix}
\label{close}
\end{align}
where $u^{t}=u^{(0)}+\delta u^t$, $v^{t}=v^{(0)}+\delta v^t$ with $\left(u^{(0)},v^{(0)}\right)= \left(d\cos\theta, d\sin\theta\right)$
and $\delta u^t=\delta u^{t+2\pi/\Omega},\, \delta v^t=\delta v^{t+2\pi/\Omega}$ . The pair $\left(u^{(0)},v^{(0)}\right)$ determines
the center of the adiabatic cycle, reparameterized with $d$ and $\theta$ in polar form.

When $u^{t}=v^{t}=0$, the eigen-frequencies of Hamiltonian Eq. (\ref{close}) are $a_1^{(0)}=a_2^{(0)}=\omega_{\rm DP}=
\omega_{0}-1$ (DP
degeneracy) and $a_3^{(0)}=\omega_{0}+2$. This DP degeneracy can be lifted in two ways: (a) by introducing $\left(u^{(0)},
v^{(0)}\right)\neq (0,0)$; (b) by coupling the system to the leads. In the latter case, the eigen-modes turn into resonant modes. 
The resonant frequencies are the real parts of the poles of the $S$-matrix which are identified with the complex eigenvalues 
$\left\{a_n\right\}$ of the effective Hamiltonian $H_{\rm eff}^{(0)}$, see Eqs. (\ref{CMT},\ref{St}) in the absence of modulation. 
Their corresponding imaginary part is the resonance line-width which describes the decay rate of these modes to the reservoirs. 
In other words the degenerate eigenmodes $a_{1,2}^{(0)}$ move apart from one-another and turn to $a_{1}\neq a_{2}$ whenever 
the above two mechanisms are in effect. The degree of repulsion between $(a_1,a_2)$ is controlled by the interplay 
of the proximity of the adiabatic cycle to the DP and the coupling with the leads.

{\it Effects of DP in Pumped Thermal Radiation--} In the presence of small time-periodic variations $(\delta v^t,\delta u^t)$, the 
Green's function $G_{\rm eff}=(H_{\rm eff}-\omega I_{N_s})^{-1}$ which appears in the evaluation of $S^t$ (and therefore in $Q(\omega)$), 
involves the instantaneous effective Hamiltonian $H_{\rm eff}=H_{\rm eff}^{(0)}+\Delta^t$ where $\Delta^t={\rm diag}\left(\delta 
v^t, 0, \delta u^t\right)$.  We expand $G_{\rm eff}$ in a series keeping terms up to first order in $\Delta^t$ i.e. $G_{\rm eff}=
G_{\rm eff}^{(0)}-G_{\rm eff}^{(0)} \Delta^t G_{\rm eff}^{(0)} + \cdots$ where $G_{\rm eff}^{(0)}=\left(H_{\rm eff}^{(0)}-\omega I_{N_s}
\right)^{-1}$. We can make further progress by representing $G_{\rm eff}$ in the bi-orthogonal basis $\left\{\left(a_{i}\right|, 
\left|a_{i}\right)\right\} (i=1,2,3)$ of $H_{\rm eff}^{(0)}$ where $G_{\rm eff}^{(0)}$ is diagonal. The reflection coefficient $R^t$ 
and phase $\alpha^t$ are evaluated (up to first order in $\delta u^t, \delta v^t$) by direct substitution of $G_{\rm eff}$ into 
Eq. (\ref{St}). Finally, the adiabatic thermal radiation energy $Q$ is calculated using Eq. (\ref{Q_general}). Specifically, 
using Eqs.~(\ref{param}, \ref{Q_general},  \ref{St}, \ref{Ra}), we find that
\begin{widetext}
\begin{align}
Q\left(\omega\right) \equiv \frac{\partial}{\partial\omega}P\left(\omega,d,\{w_n\}\right) & = 
\lim_{{\mathcal A}\rightarrow0}{\imath\over {\mathcal A}}\frac{\partial}{\partial\omega}\left\{ \int_{0}^{2\pi/\Omega}dt\mathrm{Re}\left[\left(1+\mathrm{Tr}A_{11}\right)\mathrm{Tr}\left(\tilde{\triangle}^{t}A_{11}\right)^{*}\right]
\frac{d\mathrm{Tr}\left(\frac{\tilde{\triangle}^{t}A_{22}}{1+\mathrm{Tr}A_{22}}-\frac{\tilde{\triangle}^{t}A_{11}}
{1+\mathrm{Tr}A_{11}}\right)}{dt}\right\} ,
\label{Qg}
\end{align}
\end{widetext}
where $A^{ij}=\imath\frac{W^{T}\left|a_{i}\right)\left(a_{j}\right|W}{\sqrt{a_{i}-\omega}
\sqrt{a_{j}-\omega}}$ and $\tilde{\triangle}_{ij}^{t}=\frac{\left(a_{i}\right|\triangle^{t}\left|a_{j}\right)}{\sqrt{a_{i}-\omega}\sqrt{a_{j}
-\omega}}$. The implicit subscript (not shown) of matrix $A^{ij}$ denotes the entry while the trace operation is with respect to the dummy variable, e.g., $\mathrm{Tr}\left(\tilde{\triangle}^{t}A_{11}\right)=\sum_{i,j}\tilde{\triangle}_{ij}^{t}A_{11}^{ji}$. 

Direct substitution of Eq. (\ref{Qg}) in Eq. (\ref{Ibar}) allows us to perform an integration by parts and express the thermal energy 
flux per pumping area ${\bar {\mathcal I}}$ as
\begin{equation}
\label{part}
\bar{\mathcal I} =-{\Omega \over 2\pi} \int{d\omega\over 2\pi} \frac{\partial \left[\hbar \omega \Theta_T(\omega)\right]}{\partial \omega} 
P(\omega, d, \{w_n\})
\end{equation}
where for near-resonant thermal radiation the boundary contributions (associated with the integration by part) are neglected. 
Using the residue theorem we get
\begin{equation}
\label{final1}
\bar{\mathcal I} \propto \sum_n{\rm Res} \left[{\partial f(\omega)\over \partial \omega}P(\omega);\omega_n\right];\quad {\mathcal I}m
\left(\omega_n\right)>0
\end{equation}
where $f(\omega)\equiv\hbar \omega \Theta(\omega, T)$ and $\omega_n$ are the poles
of $P(\omega)$ (note that $f(\omega)$ is an analytic function). In the case of near-field resonant thermal transport, these poles
can be associated with the poles of the scattering matrix Eq. (\ref{St}). They originate from the DP degeneracies of the isolated 
system once they move to the complex frequency plane due to coupling with the leads.

In fact, the singular behavior of $R^t$ and $d\alpha^t/dt$ will be significant in the presence of a diabolic point degeneracy. In
this case a further progress can be made for the evaluation of $\bar{\mathcal I}$. Specifically, 
\begin{equation}
\label{final2}
\bar{\mathcal I} \propto {\sum_n}^{\prime}{\rm Res} \left[P(\omega);\omega_n\right]
\end{equation}
where the summation is restricted to poles near the DP.

\begin{figure}
\includegraphics[width=1\columnwidth,keepaspectratio,clip]{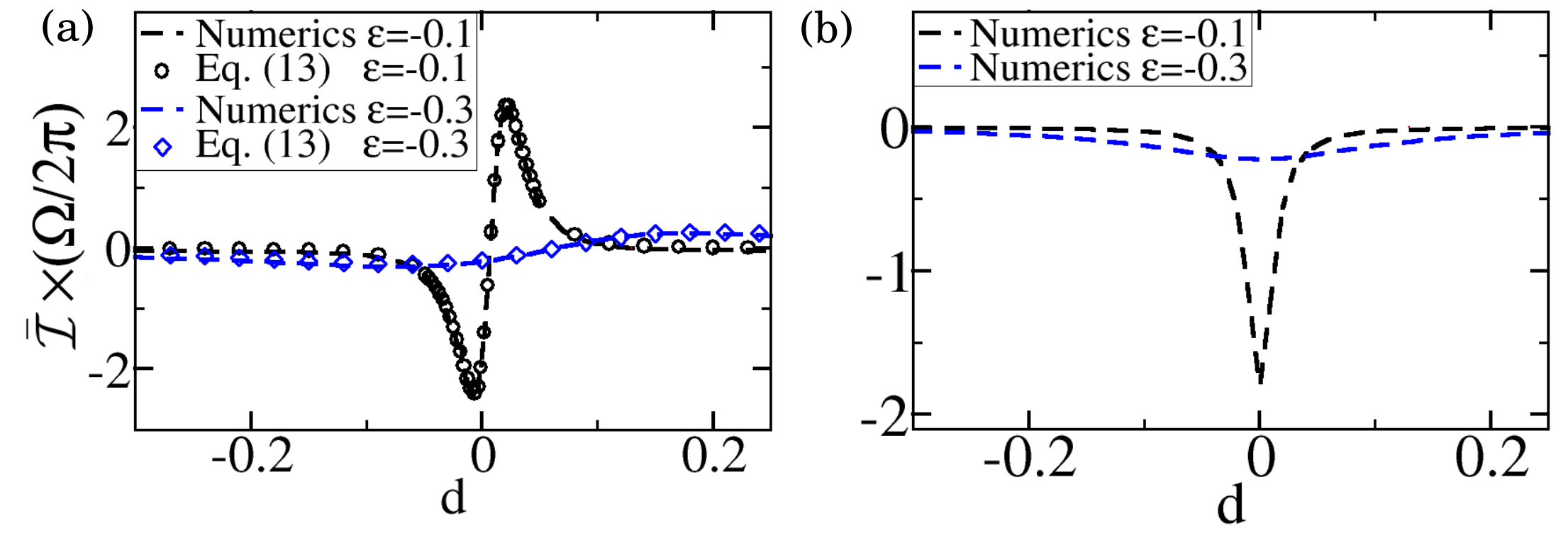}
\caption{Numerical evaluation (dashed lines) of rescaled total radiative (time-averaged) thermal energy flux $\bar{\mathcal I}\times {2\pi\over 
\Omega}$ versus the control parameter $d$ for driving angles (a) $\theta = 45^{\circ}$ and (b) $\theta = 135^{\circ}$. The coupling $\epsilon$ 
between the leads and the system is indicated in the inset of the figures. In (a) we also report the theoretical result (symbols) of Eq.~(
\ref{Q_avg_appro}). Other parameters are $k_BT=0.8$ and $\omega_{0}=3$ (in units of coupling strength).
\label{fig2} }
\end{figure}

{\it Examples of Adiabatic Pumps --} The above theoretical considerations can be directly tested using the CMT system of Eq. (\ref{close}). 
We chose $\theta=45^0$, $\delta u^t=r\cos\Omega t$ and $\delta v^{t}=r\sin \Omega t$. Furthermore we assume (left and right) 
tight-binding leads with dispersion $\omega=\omega_0-2\cos k$ that are coupled to the scattering target with the same coupling constant 
$w_L=w_R=\epsilon$. We get the following expression for the total radiative (time-averaged) thermal energy flux $\bar{\mathcal I}$ per 
pumping area (see Appendix \cite{supl})
\begin{align}
\tilde{\mathcal I} & \approx{\Omega\over 2\pi}\left.\frac{\partial f}
{\partial k}\right|_{\omega=\omega_{\rm DP}}\frac{4\sqrt{3}\varepsilon^{4}\left(\varepsilon^{2}-\sqrt{2}d\right)}
{\left[\left(\varepsilon^{2}-\sqrt{2}d\right)^{2}+12\varepsilon^{4}\right]^{2}}.
\label{Q_avg_appro}
\end{align}
In Fig. \ref{fig2}a we report Eq. (\ref{Q_avg_appro}), together with the numerical calculations using Eq. (\ref{Ibar},\ref{Q_general}). We have 
performed similar calculations for different angles $\theta$ in order to validate the general features of Eq. (\ref{Q_avg_appro}). Specifically, 
in Fig. \ref{fig2}b, we report the behavior of  $\bar{\mathcal I}$ versus the control parameter $d$ for $\theta=135^0$. In both cases we have found that 
$\bar{\mathcal I}$ diminishes for large values of the control parameter $d$ (see also Eq. (\ref{Q_avg_appro})). In this limit, the Hamiltonian 
$H_0$ of the isolated system does not support a DP. As $|d|\rightarrow 0$, the DP is re-established, leading to an enhanced radiative 
thermal energy flux. This enhancement can be further boosted by decreasing the coupling $\epsilon$ between the system and the leads.
As discussed previously, the coupling shifts the degenerate levels to the complex plane, thus lifting the DP degeneracy and therefore 
deteriorating the performance of the pump. We point out that typically, the extrema of $\bar{\mathcal I}$ is in the {\it vicinity} of the DP 
corresponding to $H_0(d=0;r=0)$ (see other cases at the Appendix \cite{supl}). This is due to the reminiscent effects of the coupling, 
which results in differences between $H_0$ and $H_{\rm eff}$(see Eq. (\ref{CMT})). Nevertheless, there can be appropriate choices of the 
control parameters $d, \epsilon$ for which these reminiscent effects disappears completely and the extrema of $\bar{\mathcal I}$ occurs 
at $d=0$, see Fig. \ref{fig2}b.

{\it Adiabatic pumping using Circuits--}
An experimental demonstration of the effects of DPs on the adiabatically pumped thermal radiation, can be achieved using the electrical 
circuit shown in Fig. \ref{fig1}b. The system consists of a chain of three LC resonators with resonant frequency $\omega_0=1/\sqrt{L C}$. The 
resonators are capacitively coupled along the chain via capacitances $\lambda C$. Moreover, the first and last resonators are coupled 
through a mutual inductance $M=\mu L$, leading to system coupling similar to Eq.~\ref{close}. The circuit is connected in series to two terminal capacitors $C_e=\epsilon C$ which we initially consider as
grounded, i.e. $V_{L(R)}=0$. For concreteness, we will set $\omega_0/2\pi =1GHz$, $\lambda=0.2$, $\epsilon=0.1$, and $z_0=\omega_0 L=1/
\omega_0C=70$ Ohm. When $\mu=\mu_{\rm DP}\approx 0.13$, the eigen-frequencies of the grounded system demonstrate a DP degeneracy 
(see Supplement \cite{supl}).

We now turn the system into a scattering set-up by coupling each terminal capacitor $C_e$ to independent reservoirs at the same temperature 
$T$. These reservoirs are represented by a model for bandwidth limited Thevenin equivalent TEM transmission lines with characteristic 
impedance $Z_0=50$ Ohms. The noise sources $V_n$ are  synthesized such that $\langle V_n(\omega)V^*_m(\omega^{\prime})\rangle=
\frac{2Z_0}{\pi}\Phi(\omega)\delta(\omega-\omega^{\prime})\delta_{nm}$ where $\Phi(\omega) = k_BT\Theta(\omega)$. For demonstration 
purposes, we set $\Theta(\omega)=\sqrt{1-\left(\omega\over 2\omega_c\right)^2}$, with $\omega_c\approx 0.47$ GHz.  

The pumping scheme is chosen to always enclose the DP when $\mu=\mu_{DP}$. Specifically, we consider
a periodic modulation of the capacitances at left and right resonators such that $C_1(t) = C\left( 1 + r\sin(\Omega t)\right), \quad   C_2(t) = C\left( 1 + r\cos(\Omega t)\right)$. 

\begin{figure}
    \includegraphics[width=1\columnwidth,keepaspectratio,clip]{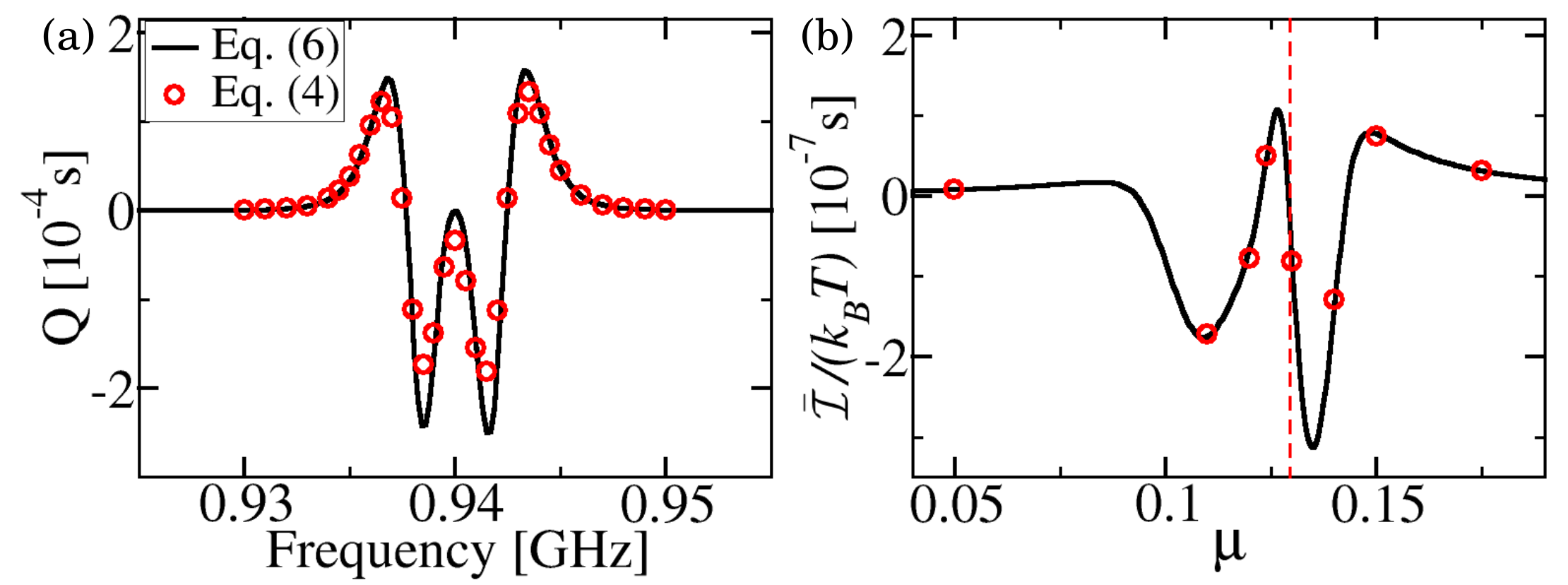}
    \caption{(a) Total pumped energy density $Q(\omega)$ versus frequency for a typical value of the rescaled mutual
inductance coupling at DP, $\mu=\mu_{\rm DP}$. The line is the numerical evaluation of $Q(\omega)$ using Eq. (\ref{Q_general}). The 
symbols are results from a direct time-domain simulation of net power at the left terminal. \cite{NGspice} (b) Total (averaged) pumped radiative energy current 
$\bar{\cal I}$ versus the mutual inductance $\mu$. The highest values of $\bar{\cal I}$ are reached at the proximity of the DP (red dashed line). 
}
  \label{fig3}
\end{figure}

Next we inject into the circuit uncorrelated incoming waves of the same frequency $\omega$ and power $P_s=V_s^2/(8Z_0)$ from the 
left (L) and right (R) reservoir. The average (over a cycle) net power flowing through the node $L(R$) is obtained from the voltage 
$v_{L (R)}(t)$ and current $i_{L (R)}(t)$ sampled at the respective 
node $L (R)$ (see Supplement \cite{supl}).
Specifically, $Q(\omega)$ is evaluated using 
Eq. (\ref{Ibar}) where the time-dependent energy current is ${\cal I}_{L(R)}(t,\omega)=v_{L(R)}(t,\omega)i_{L(R)}(t,\omega)/P_s$. In 
our simulations, we made sure the system reached a stationary state before the evaluation. The results from the time domain simulations are 
shown in Fig. \ref{fig3}a together with the outcome from the instantaneous $S^t$-matrix approach, see Eq. (\ref{Q_general}).\cite{NGspice} In the 
latter case we have extracted the instantaneous reflectance $R^t$ and reflection phase $\alpha^t$ using a standard scattering approach 
(see Supplement \cite{supl}) \cite{LSEK12}.

Having at our disposal the total radiative energy density $Q(\omega)$ for the circuit set-up, we are now able to incorporate $\Phi(\omega)$ for the total radiative 
thermal energy flux (per pumping area) passing through the system, $\bar{\cal I}$ versus $\mu$. In Fig. \ref{fig3}b we report our findings using the instantaneous $S^t$
matrix and the direct time-domain approaches. The data nicely demonstrates the enhancement in ${\cal I}$ due to the presence of the diabolic point as 
expected from the predictions of CMT.

{\it Conclusions -- } We have introduced the concept of adiabatic thermal radiation pumps as a means to manage the direction of net 
radiative energy current for bodies in equilibrium. We addressed this problem by appropriately adopting, and establishing in the framework 
of resonant near-field thermal radiation, an instantaneous scattering matrix formalism borrowed from mesoscopic condensed matter. 
Using this tool, we highlighted the importance of wave interference effects in the field of thermal photonics for thermal radiation management. 
As an example, we demonstrated the impact of diabolic point spectral singularities in such framework. Our results have been tested 
against realistic simulations using electronic circuits, with the techniques directly applicable to systems where temporal control of 
resonant coupling is possible. An exciting application of our proposal might involve tunable superconducting resonators~\cite{AKD16,BHG16} 
which will enable new forms of superconducting Q-bit manipulation. A future interesting direction is the study of the full counting statistics
for thermal radiation. It will also be interesting to extend this study to other types of spectral singularities, like exceptional points 
\cite{CLEK17}. These questions will be addressed in a subsequent publication.

{\it Acknowledgements --}(HL, LJFA, TK) acknowledge partial support by an ONR Grant No. N00014-16-1-2803, by a DARPA NLM program 
via Grant No. HR00111820042, by an AFOSR Grant No. FA 9550-14-1-0037, and by a NSF Grant No. EFMA-1641109. (BS) acknowledges
the hospitality of Wesleyan Univ. where this work has been performed.



\begin{thebibliography}{1}
\bibitem{note2}Note that $Q
\left(\omega\right)$ does not depend on the position $x_0$ where the flux is evaluated because on average no net energy is 
accumulated in the set-up.
\end{thebibliography}
\end{document}


\title{Adiabatic Thermal Radiation Pumps for Thermal Photonics}
\author{Huanan Li$^{1,2}$\footnotemark[1], Lucas J. Fern\'andez-Alc\'azar$^{1}$\footnotemark[1], Fred Ellis$^1$, Boris Shapiro$^3$, Tsampikos Kottos$^1$} 
\affiliation{$^1$Wave Transport in Complex Systems Lab, Department of Physics, Wesleyan University, Middletown, CT-06459, USA\\
$^2$Photonics Initiative, Advanced Science Research Center, CUNY, NY 10031, USA\\
$^3$Technion - Israel Institute of Technology, Technion City, Haifa 32000, Israel}
\date{\today}

\maketitle

\footnotetext[1]{These authors contributed equally to the results of this paper}
\begin{center}
\textbf{\large Supplemental Materials}
\end{center}

\setcounter{equation}{0}
\setcounter{figure}{0}
\setcounter{table}{0}
\setcounter{page}{1}
\makeatletter
\renewcommand{\theequation}{S\arabic{equation}}
\renewcommand{\thefigure}{S\arabic{figure}}
\renewcommand{\bibnumfmt}[1]{[S#1]}
\renewcommand{\citenumfont}[1]{S#1}

\section{Derivation of Eq. (13)}

For ``the examples of adiabatic pump'' in the main text, we assume
the driving angle $\theta=45^{\circ}$ and consider that the system
is coupled with two identical tight-binding leads with equal coupling
strength $\varepsilon$. The tight-binding leads supports a dispersion
$\omega=\omega_{0}-2\cos k$ in unit of coupling. In this case, the
renormalization term $\Lambda$ appearing in Eq. (1) of the main text
take the specific form $\Lambda=-\frac{1}{2}\cot\left(k\right)WW^{T}$
with the coupling matrix $W$ being

\begin{equation}
W=\sqrt{2\sin k}\begin{bmatrix}\varepsilon & 0\\
0 & 0\\
0 & \varepsilon
\end{bmatrix}.
\end{equation}
Notice that the free space dispersion is a special case for which
$k\approx\pi/2$ (wide-band approximation) and thus $\Lambda=0$.
On the other hand, the instantaneous effective Hamiltonian $H_{\mathrm{eff}}=H_{\mathrm{eff}}^{(0)}+\triangle^{t}$
reads 
\begin{align}
H_{\mathrm{eff}}^{(0)} & =\begin{bmatrix}\omega_{0}+\varepsilon_{k}\left(d\right) & -1 & 1\\
-1 & \omega_{0} & -1\\
1 & -1 & \omega_{0}+\varepsilon_{k}\left(d\right)
\end{bmatrix},
\end{align}
and $\triangle^{t}=\mathrm{diag}\left(\delta v^{t},0,\delta u^{t}\right)$ where $\varepsilon_{k}\left(d\right)=-\varepsilon^{2}e^{\imath k}+\frac{d}{\sqrt{2}}$,
$\delta v^{t}=r\sin\Omega t$ and $\delta u^{t}=r\cos\Omega t$. The
eigenvalue system of $H_{\mathrm{eff}}^{(0)}=H_{\mathrm{eff}}^{(0),T}$,
i.e., $H_{\mathrm{eff}}^{(0)}\left|a_{i}\right)=a_{i}\left|a_{i}\right),\:i=1,2,3$.,
can be given explicitly 
\begin{align}
a_{1} & =\omega_{0}-1+\varepsilon_{k}\left(d\right)\nonumber \\
a_{2} & =\omega_{0}+\frac{1}{2}\left[1+\varepsilon_{k}\left(d\right)-\sqrt{9+\varepsilon_{k}\left(d\right)\left(2+\varepsilon_{k}\left(d\right)\right)}\right]\\
a_{3} & =\omega_{0}+\frac{1}{2}\left[1+\varepsilon_{k}\left(d\right)+\sqrt{9+\varepsilon_{k}\left(d\right)\left(2+\varepsilon_{k}\left(d\right)\right)}\right]\nonumber 
\end{align}
and 
\begin{align}
\left|a_{1}\right) & =\frac{1}{\sqrt{2}}\begin{bmatrix}-1 & 0 & 1\end{bmatrix}^{T}\nonumber \\
\left|a_{2}\right) & =\frac{1}{\sqrt{2+\left(a_{3}-\omega_{0}\right)^{2}}}\begin{bmatrix}1 & a_{3}-\omega_{0} & 1\end{bmatrix}^{T}\\
\left|a_{3}\right) & =\frac{1}{\sqrt{2+\left(a_{2}-\omega_{0}\right)^{2}}}\begin{bmatrix}1 & a_{2}-\omega_{0} & 1\end{bmatrix}^{T}.\nonumber 
\end{align}
Due to the transposition symmetry of the matrix $H_{\mathrm{eff}}^{(0)}$,
$\left(a_{i}\right|\equiv\left|a_{i}\right)^{T}$ together with $\left|a_{i}\right)$
constitute a complete biorthonormal basis. Correspondingly, using
Eq. (9), we have the radiative energy density

\begin{align}
Q\left(\omega\right) & =\frac{\partial}{\partial\omega}P\left(k,d,\varepsilon\right),
\end{align}
where 
\begin{widetext}
\begin{align}
P\left(k,d,\varepsilon\right) & =\frac{\frac{d\delta\omega}{dk}\frac{\varepsilon^{6}\left(2-\varepsilon^{2}\right)}{\left(1-\varepsilon^{2}\right)^{4}}\left(4-\delta\omega^{2}\right)\delta\omega\left(\delta\omega+1\right)^{2}\left(\delta\omega-\omega_{5}\right)\left(\delta\omega-\omega_{6}\right)}{\left(\delta\omega-\omega_{1}\right)^{2}\left(\delta\omega-\omega_{1}^{*}\right)^{2}\left(\delta\omega-\omega_{2}\right)^{2}\left(\delta\omega-\omega_{2}^{*}\right)^{2}\left(\delta\omega-\omega_{3}\right)^{2}\left(\delta\omega-\omega_{4}\right)^{2}}\label{eq: S_P(k,d)}
\end{align}
\end{widetext}
with the frequency detuning $\delta\omega\equiv\omega-\omega_{0}=-2\cos k$
and in the limit of $d\sim O\left(\varepsilon^{2}\right)\rightarrow0$
\begin{align}
\omega_{1} & \approx-1+\frac{\sqrt{2}}{2}d+\frac{1}{2}\left(-1+\imath\sqrt{3}\right)\varepsilon^{2}\nonumber \\
\omega_{2} & \approx-1+\frac{\sqrt{2}}{6}d+\frac{1}{6}\left(-1+\imath\sqrt{3}\right)\varepsilon^{2}\nonumber \\
\omega_{3} & \approx\omega_{4}\approx2+\frac{\sqrt{2}}{3}d+\frac{2}{3}\varepsilon^{2}\\
\omega_{5} & \approx1+\frac{\sqrt{2}}{4}d+\frac{1}{4}\varepsilon^{2}\nonumber \\
\omega_{6} & \approx-1+\frac{\sqrt{2}}{4}d-\frac{1}{4}\varepsilon^{2}.\nonumber 
\end{align}

\begin{figure}
    \includegraphics[width=0.8\textwidth ]{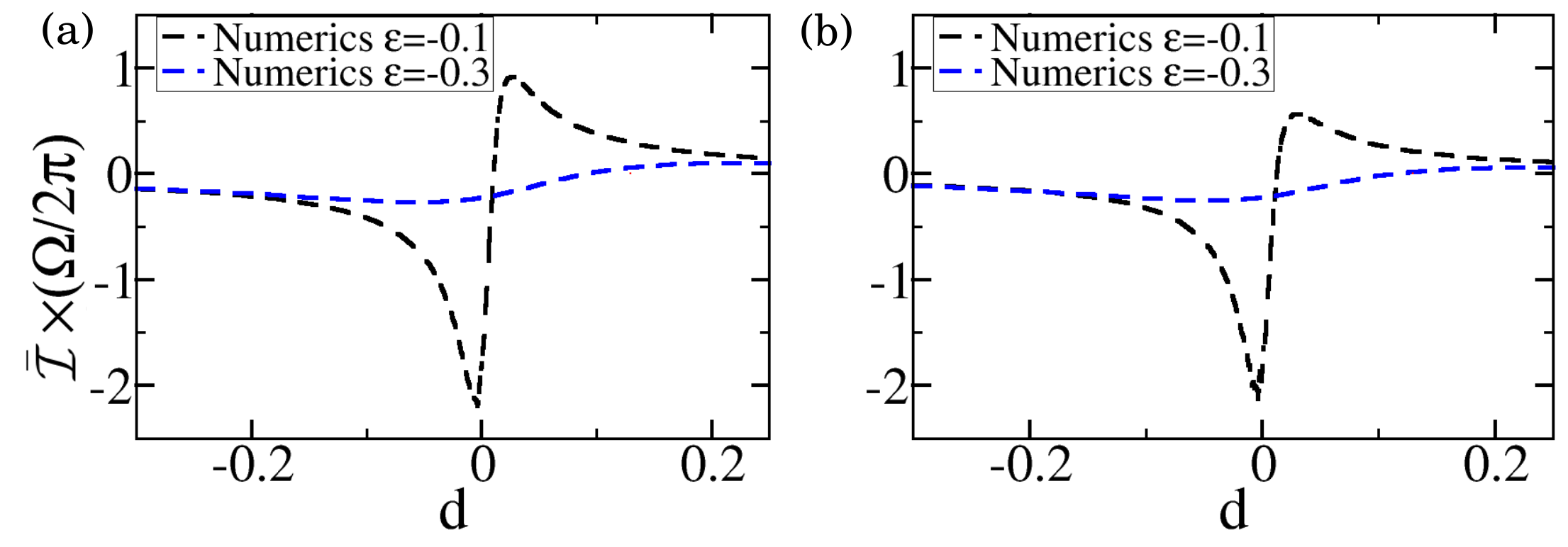}
\caption{Numerical calculation of $\bar{\mathcal{I}}\frac{2\pi}{\Omega}$ versus
$d$ for two different coupling strengths $\varepsilon=-0.1,-0.3$
when (a) $\theta=0^{\circ}$, (b) $\theta=100^{\circ}$. Other parameters
are $\beta=0.8$ and $\omega_{0}=3$.\label{fig: I_other_theta} }
\end{figure}

For the thermal energy flux per pumping area $\bar{\mathcal{I}}$,
Eq. (10) in the main text can be rewritten as 
\begin{align}
\bar{\mathcal{I}} & =-\frac{\Omega}{2\pi}\int_{0}^{\pi}\frac{dk}{2\pi}P\left(k,d,\varepsilon\right)\frac{\partial}{\partial k}f\left(\omega_{0}-2\cos k\right).\label{eq: S_I}
\end{align}
The major contribution of $P\left(k,d,\varepsilon\right)$ to the
integral mainly comes from $k$ values of short range around DP satisfying
$\delta\omega\approx-1$, see Eq.~(\ref{eq: S_P(k,d)}). Therefore
we can simplify Eq.~(\ref{eq: S_I}) for $\bar{\mathcal{I}}$ to
be 
\begin{align}
\bar{\mathcal{I}} & \approx-\frac{\Omega}{2\pi}\left.\frac{\partial f}{\partial k}\right|_{_{\delta\omega=-1}}\frac{\varepsilon^{6}\left(2-\varepsilon^{2}\right)}{\left(1-\varepsilon^{2}\right)^{4}}\frac{3\left(1+\omega_{5}\right)}{\left(1+\omega_{3}\right)^{2}\left(1+\omega_{4}\right)^{2}}I_{R}\nonumber \\
I_{R} & \equiv\int_{-2}^{2}\frac{d\left(\delta\omega\right)}{2\pi}\frac{\left(\delta\omega+1\right)^{2}\left(\delta\omega-\omega_{6}\right)}{\left(\delta\omega-\omega_{1}\right)^{2}\left(\delta\omega-\omega_{1}^{*}\right)^{2}\left(\delta\omega-\omega_{2}\right)^{2}\left(\delta\omega-\omega_{2}^{*}\right)^{2}}.\label{eq: I_appro}
\end{align}
After extending the integration range of $\delta\omega$ to the whole
real axis and using residue theorem, we can evaluate $I_{R}$ in Eq.~(\ref{eq: I_appro})
approximately. Finally, we have 
\begin{align}
\bar{\mathcal{I}} & \approx\frac{\Omega}{2\pi}\left.\frac{\partial f}{\partial k}\right|_{_{\delta\omega=-1}}\frac{4\sqrt{3}\varepsilon^{4}\left(\varepsilon^{2}-\sqrt{2}d\right)}{\left[\left(\varepsilon^{2}-\sqrt{2}d\right)^{2}+12\varepsilon^{4}\right]^{2}}.\label{eq: Q_avg_appro}
\end{align}
Here, we also report the numerical calculations of $\bar{\mathcal{I}}\frac{2\pi}{\Omega}$
versus $d$ for several different angles $\theta=0^{\circ},90^{\circ}$
in addition to what we show in the main text. As shown in Fig.~\ref{fig: I_other_theta},
the extrema is also in the vicinity of the DP similar to the scenario
$\theta=45^{\circ}$ shown in Fig. 2(a) of the main text.

\section{Eigenfrequencies of the electrical circuit}
Let us consider that capacitors $C_e$ are grounded, then $V_{L(R)}=0$.
At steady state, voltages at nodes $i=1,c,2$ (see Fig. 1b) are given by $v_i(t)=\mathrm{Re}\left(V_i e^{\im \omega t}\right)$, 
where $\omega$ is the angular frequency.
The Kirchoff equations for the voltage amplitudes $V_1$, $V_c$, and $V_2$ 
can be written in matrix form as $A \vec{V}=0$, 
\begin{equation}
\begin{pmatrix}
{{\omega}^{2}} \tilde{C}_1 -1/\tilde{L} & -{{\omega}^{2}} C  \lambda & \mu/\tilde{L} \cr
-{{\omega}^{2}} C  \lambda & {{\omega}^{2}}\tilde{C}_c  -1/L & -{{\omega}^{2}} C  \lambda  \cr
\mu/\tilde{L} & -{{\omega}^{2}} C  \lambda & {{\omega}^{2}} \tilde{C}_2 -1/\tilde{L}
\end{pmatrix}
\begin{pmatrix}
V_1 \cr V_c \cr V_2
\end{pmatrix}
=0,
\label{eq:voltages1}
\end{equation}
where $\tilde{C}_1=C_e+C_1+C  \lambda$, 
$\tilde{C}_c=\left( C+2 C  \lambda\right)$, 
$\tilde{C}_2=C_e+C_2+C  \lambda$, and
$\frac{1}{\tilde{L}}=\frac{L}{{{L}^{2}}-{{\mu}^{2}} {{L}^{2}}}$.

The eigenfrequencies are those values of $\omega$ that give a determinant $\mathrm{det}(A)=0$. 
In absence of driving $C_1=C_2=C$, and using $\Omega=(\omega/\omega_0)^2$, $C_e/C=\epsilon$, the eigenfrequencies 
can be written as
\begin{eqnarray}
\Omega_{1,2}  &=&  \frac{2+\tilde{\epsilon}_\mu+\tilde{\mu} } {a} \pm \frac {\sqrt{
\tilde{\mu}^2 + 2\lambda \left[-\epsilon-\mu+ \mu(\tilde{\epsilon}_\mu-\epsilon)\right]  +\tilde{\epsilon}_\mu^2}}{a} \notag\\
\Omega_3  &=& -\frac{1}{ (\mu - 1) \left( 1+\lambda + \varepsilon\right) }. 
\end{eqnarray}
Here, $\tilde{\epsilon}_\mu=\epsilon+\mu(\epsilon+1)$, $\tilde{\mu}=(3+\mu)\lambda$, and $a=2(1+ \mu) \left[ (2 \epsilon+3 )\lambda + \epsilon+1\right]$.
Given $\lambda$ and $\epsilon$, the solutions $\Omega_1$ and $\Omega_3$ take the same value at a particular 
$\mu=\mu_{DP}$. For the case $\lambda=0.2$, $\epsilon=0.1$, we have $\mu_{\rm DP}\approx 0.129506$.
We note that this is not an avoided crossing, but a rare occurrence called a diabolic point.\\

\section{Scattering Parameters for the Electrical Circuit}
Now, let us consider the system depicted in Fig. 1b as a scatterer where
the voltages on the left and right nodes, $V_L$ and $V_R$, are not grounded. Instead, they are driven by 
time dependent voltage sources $v^s_{L(R)}(t)=\mathrm{Re} \left( V^s_{L(R)} e^{\im \omega t} \right)$.
We can split both the voltages at, and the currents 
passing through, the nodes $L(R)$
from the model transmission line
as a sum of two contributions,
\begin{eqnarray}
	 v_{L (R)}(t,\omega) &=& \mathrm{Re} \left[ (V^+_{L(R)} + V^-_{L(R)}) e^{\im \omega t} \right],  \notag \\
	 i_{L (R)}(t,\omega) &=& \frac{1}{Z_0}\mathrm{Re}\left[(V^+_{L(R)} - V^-_{L(R)})e^{\im \omega t}\right].
\end{eqnarray}
Here $V^+_L=V^s_{L}/2$ and $V^-_R=V^s_{R}/2$ are the complex voltage amplitudes of the incoming waves from left and right Thevenin equivalent sources;
and $V^-_L$ and $V^+_R$ are the voltage amplitudes of the outgoing scattered waves (see green arrows in Fig. 1b).
In the particular case in which $V_R^-=0$, $ V^-_L$ ($V^+_R$) is the amplitude of the reflected (transmitted) wave. Then, 
 $r_L=\frac{2 V_L^-}{V_s}$  ($t_{R,L}=\frac{2 V_R^+}{V_s}$) is the reflection (transmission) coefficient.
 Similarly, when $V^+_L=0$, the reflection (transmission) coefficient is $r_R=\frac{2 V_R^+}{V_s}$ ($t_{L,R}=\frac{2 V_L^-}{V_s}$).